\documentclass[pre,amsmath,amssymb,twocolumn,superscriptaddress,notitlepage]{revtex4-1}
\usepackage{bm}
\usepackage{amsfonts}
\usepackage{graphicx}
\usepackage{color}
\usepackage{siunitx}
\usepackage[version=4]{mhchem}
\usepackage{hyperref}
\hypersetup{
    colorlinks=true,
    linkcolor=blue,
    filecolor=magenta,      
    urlcolor=cyan,
}

\newcommand{\nuphys}{Department of Physics and Astronomy, Northwestern University, Evanston, IL 60208, USA}
\newcommand{\buchem}{Department of Chemistry, University of Massachusetts Boston, Boston, MA 02125, USA}
\newcommand{\buphys}{Department of Physics, University of Massachusetts Boston, Boston, Massachusetts 02125, USA}
\newcommand{\buqns}{Center for Quantum and Nonequilibrium Systems, University of Massachusetts Boston, Boston, MA 02125, USA}
\newcommand{\nico}{Northwestern Institute on Complex Systems, Northwestern University, Evanston, IL 60208, USA}

\begin{document}
\title{Non-normality and non-monotonic dynamics in complex reaction networks}
\author{Zachary G. Nicolaou}
\affiliation{\nuphys}
\author{Takashi Nishikawa}
\affiliation{\nuphys}
\affiliation{\nico}
\author{Schuyler B. Nicholson}
\affiliation{\buchem}
\affiliation{\buqns}
\author{Jason R. Green}
\affiliation{\buchem}
\affiliation{\buqns}
\affiliation{\buphys}
\author{Adilson E. Motter}
\affiliation{\nuphys}
\affiliation{\nico}

\begin{abstract}
Complex chemical reaction networks, which underlie many industrial and biological processes, often exhibit non-monotonic changes in chemical species concentrations, typically described using nonlinear models.  Such non-monotonic dynamics are in principle possible even in linear models if the matrices defining the models are non-normal, as characterized by a necessarily non-orthogonal set of eigenvectors.  However, the extent to which non-normality is responsible for non-monotonic behavior remains an open question.  Here, using a master equation to model the reaction dynamics, we derive a general condition for observing non-monotonic dynamics of individual species, establishing that non-normality promotes non-monotonicity but is not a requirement for it.  In contrast, we show that non-normality is a requirement for non-monotonic dynamics to be observed in the R\'{e}nyi entropy.  Using hydrogen combustion as an example application, we demonstrate that non-monotonic dynamics under experimental conditions are supported by a linear chain of connected components, in contrast with the dominance of a single giant component observed in typical random reaction networks.  The exact linearity of the master equation enables development of rigorous theory and simulations for dynamical networks of unprecedented size (approaching $10^5$ dynamical variables, even for a network of only $20$ reactions and involving less than $100$ atoms).  Our conclusions are expected to hold for other combustion processes, and the general theory we develop is applicable to all chemical reaction networks, including biological ones.
\end{abstract}

\maketitle

Matrix non-normality is perhaps best known for its role in a counter-intuitive form of nonlinear instability~\cite{Trefethen:1993}.  Even when a fixed point is linearly stable in a nonlinear system described by ordinary differential equations, if the corresponding Jacobian matrix is non-normal, a small but finite perturbation can transiently grow beyond the validity of the linear approximation and enter into the nonlinear regime, preventing the perturbation from decaying to zero.  The discovery of this phenomenon has led to a thorough study of the spectral properties of non-normal matrices in the context of transient dynamics~\cite{2005_Trefethen}; it has also inspired recent works on implications of non-normality for network and spatiotemporal dynamics~\cite{Neubert:1997,Hennequin:2012,Tang:2014,Asllani:2018,Asllani:2018b,Nicoletti:2019,Baggio:2020,Tarnowski:2020,Johnson:2020,Biancalani:2017,Maini:2019,Klika:2017,Nishikawa:2006a,Nishikawa:2006b,Ravoori:2011}.  Given the common perception that linear dynamics are fully understood, the possibility of such transient growth offers interesting alternative interpretations for behavior usually attributed to nonlinearity, such as ignition dynamics in combustion and temporary activation of biochemical signals.

In network systems, however, even at the level of linear dynamics, fundamental questions remain open concerning such transient growth---or, more generally, non-monotonic dynamics.  How prevalent is non-normality in dynamical networks and how often does it lead to non-monotonic dynamics?  While non-normality is known to be widespread among matrices encoding network structures~\cite{Asllani:2018}, the question is open for matrices representing dynamical interactions, which have direct implications on non-monotonic dynamics.  Since non-monotonicity could be observed for one variable but not for others within the same system, how can we determine from the initial conditions whether a given variable will show non-monotonic behavior?  Beyond the known tendency of non-normality to be correlated with local and global directionality of the network~\cite{Asllani:2018,Johnson:2020,Hennequin:2012,Nishikawa:2006b,Ravoori:2011}, what other connectivity structures have implications for non-normality and/or non-monotonic dynamics?

\begin{figure*}[tp]
\includegraphics[width=\textwidth]{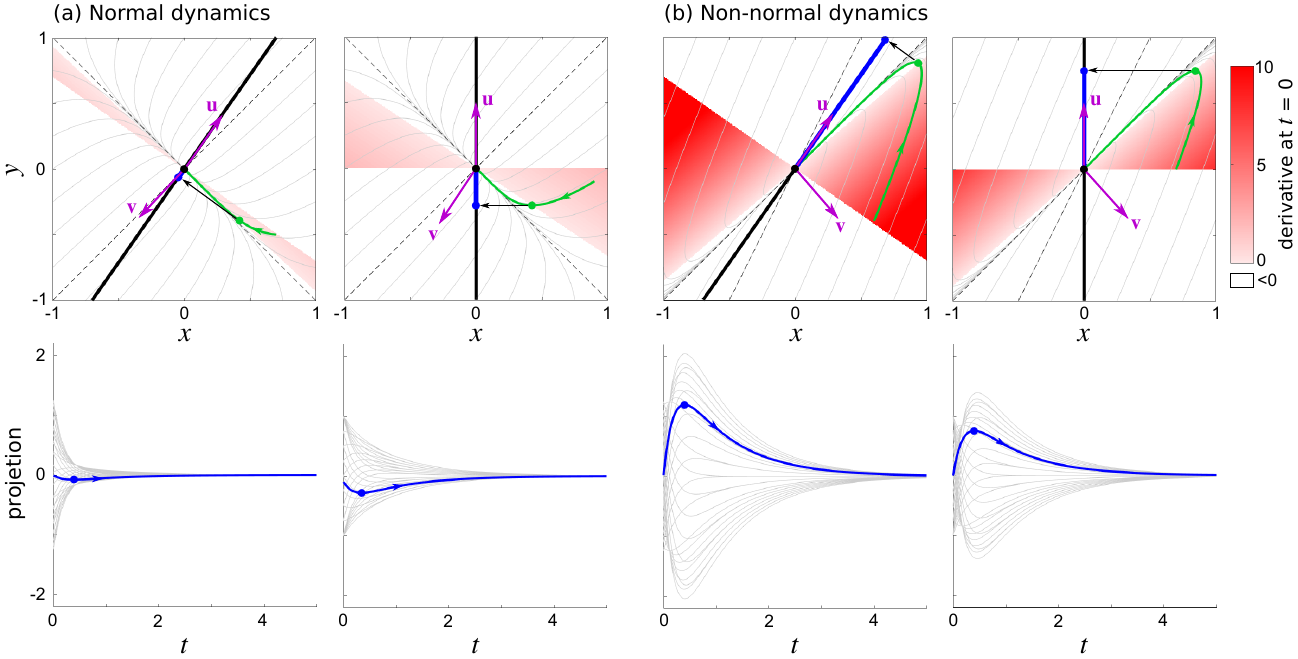}
\vspace{-5mm}
\caption{Non-monotonicity of normal and non-normal dynamics.  In each column, the top panel shows trajectories of a two-dimensional linear system (gray curves in the background and one curve highlighted in green), while the bottom one shows the projection of these trajectories onto a one-dimensional subspace (black solid line in the top panel) parallel to a given vector $\mathbf{u}$ (purple arrow) as a function of time.  In the top panel, the dashed lines indicate the two eigenvectors of the system, and the red color intensity at each point encodes the rate at which the projection of the trajectory starting from that point initially moves away from zero (i.e., the derivative of the projection at $t=0$, multiplied by the sign of the projection).  The vector $\mathbf{v}$ shown in purple is orthogonal to a boundary line of the red region (the line on which the derivative of the projection is zero at $t=0$).  (a)~Normal system, $\dot{x} = -3x-2y$, $\dot{y} = -2x-3y$, whose eigenvectors are orthogonal.  The dynamics are projected onto a line with a $55^\circ$ slope (left column) and a vertical line (right column).  The system exhibits weakly non-monotonic trajectories in both cases.  (b)~Non-normal system, $\dot{x} = 3x-4y$, $\dot{y} = 8x-9y$, whose eigenvectors are far from being orthogonal.  The trajectories are projected on to the same lines as in (a).  In contrast to the normal system in (a), this non-normal system exhibits a much more pronounced transient growth.  While the two representative projection angles are used here to illustrate the contrast between the two systems, the entire range of possible angles are shown in an animated version of this figure provided as Supplemental Material (click \href{https://northwestern.box.com/s/otn3m2cov9gi3enht3r8jh5kjo9qnv6d}{here} for the video).  The animation shows that the ranges of initial conditions and projection angles leading to non-monotonicity is wider for the non-normal system than for the normal system.
\label{fig_non_monotonic}}
\end{figure*}

In this article, we address these questions using the chemical master equation~\cite{McQuarrie:1967,1992_Gillespie}, which describes the dynamics of a chemical reaction network in terms of a time-evolving probability distribution of the network's state.  Master equations play an important role in statistical physics~\cite{Reichl:2016,Krapivsky:2010} and have been used to model physical and chemical processes in various contexts (e.g., Refs.~\cite{Seshadri:1980,Jarzynski:1997}), including processes on networks (e.g., Refs.~\cite{Pastor-Satorras:2001,Albert:2002,Hoffmann:2012}).  The chemical master equation has the advantage of being exactly linear and amenable to rigorous analysis, while still reflecting the dynamics of species concentrations, which are most often alternatively modeled by nonlinear reaction rate equations in the limit of large number of molecules.  This connection between linear and nonlinear models is possible because, when the number of molecules is large, the nonlinear dynamics in the finite-dimensional space of species concentrations can be lifted to linear dynamics in an infinite-dimensional space (which is akin to how the linear, infinite-dimensional Koopman operator can be used to study nonlinear, but finite-dimensional, dynamical systems~\cite{2019_Mezic}).  Thus, while the individual interactions between the species concentration dynamics are inherently nonlinear, the interactions between the probability dynamics of different states in the chemical master equation are strictly linear.  A key to our approach, particularly for reactions involving a finite number of molecules, is to represent these linear interactions by a directed weighted network of state-to-state transitions.

Here, we study what the structure of such a network can tell us about the chemical process it encodes.  In particular, we focus on the consequences of non-normality, strongly connected components, and reaction irreversibility underlying the local directionality of network links, examining their implications for non-monotonic dynamics.  As a representative example of real complex chemical reactions, we consider hydrogen combustion, for which a dataset is available on the experimentally determined reactions, species, and rate constants~\cite{2004_Li_Frederick,2004_Conaire_Charles,2005_Baulch,2019_Konnov}.  In the master-equation representation, the network grows rapidly in size with the number of atoms, reaching tens of thousands of nodes for fewer than a hundred atoms.  To address the computational challenges of constructing the network and solving the master equation, we developed a thresholding technique that substantially reduces the network size without significantly affecting the accuracy of system trajectory calculations.  This technique is incorporated in our open-source toolbox for automatically constructing a network from a given reaction dataset~\cite{github}.  Equipped with this toolbox, we use hydrogen combustion as a representative example to demonstrate that combustion networks are indeed non-normal, with some eigenvectors having small angles between them (and hence far from being orthogonal), leading to non-monotonic behavior in the number of molecules of intermediate chemical species.

\begin{figure*}
\includegraphics[width=\textwidth]{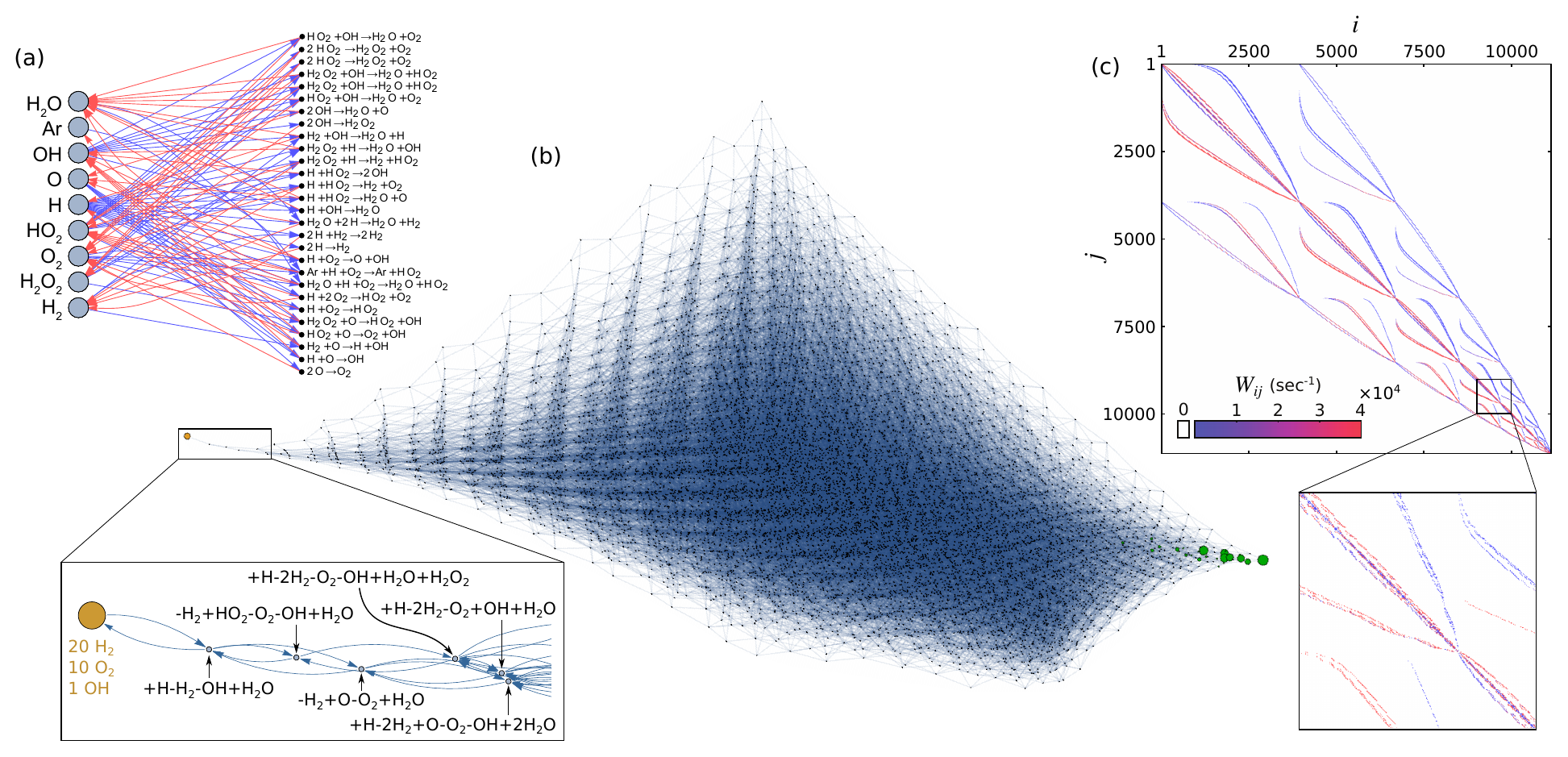}
\caption{Network representations of the \ce{H2}/\ce{O2} combustion process.  (a)~Conventional bipartite graph representation, in which the nodes representing reactions (black dots) have incoming links (blue arrows) from the nodes representing the reactant species and outgoing links (red arrows) to those representing the product species.  (b)~Master-equation representation, in which each node represents a state defined by the number of molecules of all species and a weighted link the rate of transition between two states.  The network shown encompasses the $11{,}129$ states accessible from the initial state (orange node on the left) with 20\,\ce{H2}, 10\,\ce{O2}, 1\,\ce{OH}, and 200\,\ce{Ar} molecules.  The green nodes on the right indicate the states with strictly positive probability of occupancy in the steady-state distribution of the chemical master equation, Eq.~\eqref{cme}.  All the other states are indicated by black dots, and the (reversible) transitions between the nodes are represented by links.  The inset shows a zoom-up of the initial state and its neighboring states, which are labeled with the changes in species composition relative to the initial state.  The link strengths in the inset are proportional to the transition rates $W_{ij}$.  (c)~Transition rate matrix $\mathbf{W}$ for Eq.~\eqref{cme}.  The states are indexed by the order of discovery in the state enumeration algorithm based on depth-first search, starting from the initial state $i=1$. 
\label{fig1}}
\end{figure*}

More generally, we derive a rigorous, geometrically interpretable condition on the initial probability distribution under which the time evolution of a given set of chemical species is non-monotonic.  In particular, the condition indicates that the dynamical non-monotonicity depends both on the initial distribution and the selected set of species.  The condition also shows that non-normality is not strictly required for non-monotonicity. Indeed, for a generic system, be it normal or non-normal, there is always a projection of the solution space for the master equation that leads to non-monotonicity for some initial distribution.  Thus, the key question is whether the dynamics of a given set of species are determined by such a projection.  Non-normal systems, however, distinguish themselves from their normal counterparts in that non-monotonicity is more prevalent and more pronounced (as illustrated in Fig.~\ref{fig_non_monotonic} using a simple two-dimensional system).  In addition, we establish that the relation between non-normality and non-monotonicity can be expressed in terms of the R\'{e}nyi entropy~\cite{renyi_1961}, in analogy with the known relation between (the violation of) the molecular chaos assumption in the H-theorem and non-monotonicity of the Shannon entropy~\cite{Tolman:1979}.  Furthermore, capturing the local link directionality and decomposing the network into connected components, we reveal the global directionality of the network in the form of a directed acyclic graph (DAG) linking these connected components.  Starting from an initial state most natural for the combustion process, the system traverses a linear chain of irreversible steps within the DAG structure.  By comparing with multiple classes of random networks, we conclude that the existence of this linear-chain structure must be attributed to non-random nature of the real combustion networks.

\begin{figure*}[p]
\includegraphics[height=0.835\textheight]{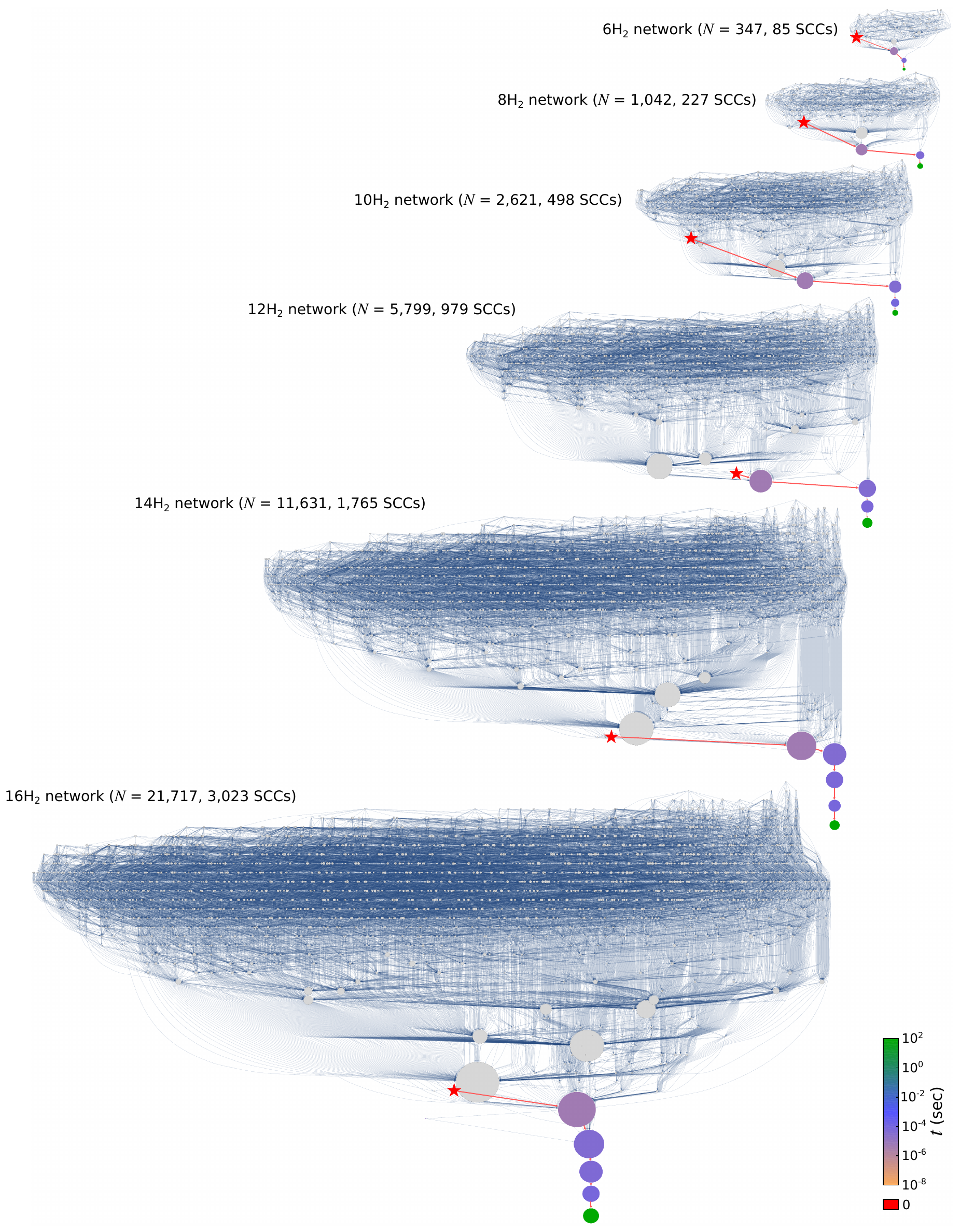}
\vspace{-3mm}
\caption{Networks of strongly connected components (SCCs), representing the structures of the weighted networks of state transitions constructed from the $6$\,\ce{H2}, $8$\,\ce{H2}, \ldots, $16$\,\ce{H2} initial states (with the numbers of \ce{O2} molecules and \ce{Ar} atoms varying proportionally).  Each node represents an SCC, a set of states connected bidirectionally by paths of state transitions. The SCCs are identified in the network constructed by enumerating states without the $\epsilon$-thresholding and then removing all transitions whose rates are much smaller than the rates of the corresponding opposite transitions (see Sec.~\ref{sec:cca} for further details).  The node size is proportional to the number of states in the corresponding SCC.  Each directed link indicates that a transition is possible from a state in one SCC to a state in the other SCC.  The red star symbol indicates the SCC containing only the initial state.  The few color-coded nodes connected by red links are the SCCs of the network obtained when applying the $\epsilon$-thresholding, with the node color encoding the time at which the maximum is observed for the probability that the system's state is in the corresponding SCC.  These SCCs are observed to form a directed linear chain in each case (when excluding SCCs with maximum probability $<10^{-3}$).
\label{fig_scc3}}
\end{figure*}

\section{Master-equation formulation}

Given a set of $N_\text{s}$ chemical species and a set of $N_\text{r}$ reactions involving them, we consider the dynamics of reactions in a mixture of these species.  The state of the system is defined by the numbers of molecules of individual species.  For a given state $i$, we denote by $m_{ik}$ the number of molecules of species $k$ in that state.  For a given reaction $n$, we denote by $\xi_{nk}$ and $\zeta_{nk}$ the numbers of molecules of species $k$ that are reactants and products, respectively.  Under this reaction, state $i$ with $m_{ik}$ molecules of each species $k$ transitions to a state with $m_{ik} - \xi_{nk} + \zeta_{nk}$ molecules of each species $k$.  This allows us to represent the system as a network of states (nodes) connected by the state-to-state transitions induced by reactions (directed links).  This representation is entirely different from (but related to) to the conventional representation of the species-reactions relation by a directed bipartite graph.  For the combustion example we describe in detail below, the bipartite representation is shown in Fig.~\ref{fig1}(a) and compared to our representation in Fig.~\ref{fig1}(b).  Assuming that the molecules are well-mixed within a fixed volume, the dynamics can be described probabilistically by the chemical master equation~\cite{1992_Gillespie}, which determines the time evolution of $P_i = P_i(t)$, the probability that the system is in state $i$ at time $t$:
\begin{equation}
\label{cme}
\frac{{\mathrm d} P_i}{{\mathrm d} t} = \sum_j W_{ij} P_j,
\end{equation}
where $W_{ij} \ge 0$ is the rate of transition from state $j$ to state $i$ given by
\begin{equation}\label{wij}
W_{ij} = \sum_n \kappa_{nj} \prod_k \frac{m_{jk}!}{ (m_{jk}-\xi_{nk})! \, \xi_{nk}!} \left(\delta_{ij'(n)}-\delta_{ij}\right),
\end{equation}
$\kappa_{nj}$ is a constant characterizing the rate at which reaction $n$ occurs when the system is in state $j$, the notation $\delta_{ij}$ is used for the Kronecker delta, and $j'(n)$ is the state to which the system transitions from state $j$ under reaction $n$.  Since the rate of a reaction depends on the numbers of molecules of the reactants but not on those of the products, $W_{ij}$ involves $\xi_{nk}$ but not $\zeta_{nk}$.  Equation~\eqref{wij} implies that the transition rates $W_{ij}$ are independent of the probabilities $P_j$, making Eq.~\eqref{cme} strictly linear.  The system trajectory starting from a specific state, which we label with $i=1$, can be determined by solving Eq.~\eqref{cme} with $P_1(0)=1$ and $P_i(0)=0$ for all $i>1$.  The dynamics can thus be regarded as a Markov process over the network of states driven by transition rates $W_{ij}$, which can be regarded as the weight of the link from node $j$ to $i$.

As a specific example of chemical reaction dynamics, we consider combustion in a gas composed of hydrogen, oxygen, and a neutral buffer of argon, under a constant-volume, constant-energy condition.  The argon buffer is included to absorb most of the energy released by the reactions, so that the evolving temperature of the gas remains within the acceptable range for the kinetic model (to be described below) while conserving energy.  In addition to the reactants and products of the overall combustion reaction, \ce{2H2 + O2 -> 2H2O}, various intermediate, short-lived, radical species are created during this process ($\ce{H}$, $\ce{O}$, $\ce{OH}$, $\ce{HO2}$, and $\ce{H2O2}$).  From a given initial state of this \ce{H2}/\ce{O2} combustion process with specified numbers of molecules of each species, the number of states that are accessible through a sequence of reactions is finite, and we denote this number by $N$.  This is because the number of each atomic species in the gas is conserved during each reaction event.  Assuming that hydrodynamic effects are negligible (i.e., the gas is well-mixed, and the kinetic energy released by reactions thermalizes quickly), each accessible state $i$ has well-defined temperature $T_i$ and pressure $p_i$ (which can vary with $i$ due to the energy released or absorbed by the reactions).  Because the temperature and pressure should not fall unrealistically low, the number of accessible states is further limited. 

\begin{figure}[tp]
\includegraphics[width=\columnwidth]{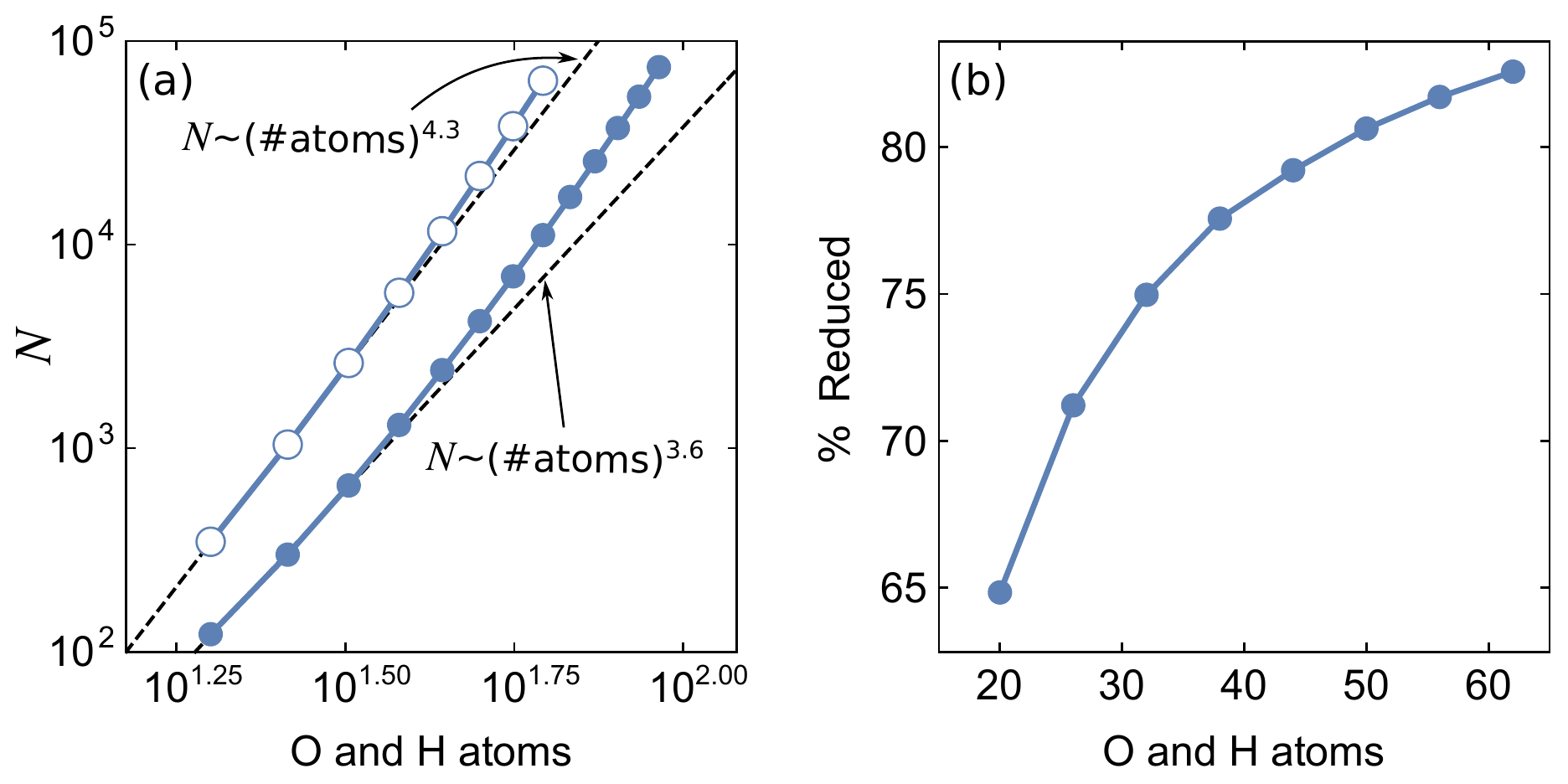}
\vspace{-6mm}
\caption{Growing size and complexity of the \ce{H2}/\ce{O2} combustion networks with increasing number of atoms.  (a)~Log-log plot of network size $N$ against the number of atoms with (solid circles) and without (open circles) the $\epsilon$-thresholding applied during state enumeration.  In each case, $N$ appears to grow faster than the power law indicated by the dashed line.  (b)~Percentage reduction in $N$ achieved by the $\epsilon$-thresholding, as a function of the number of atoms.
\label{dimension}}
\end{figure}

To map out the network of states, we implemented an algorithm based on a depth-first search to identify all states accessible from a given initial state through the network of possible transitions.  Our implementation~\cite{github} uses Cantera \cite{cantera} and can be applied to any reaction network data in the Cantera or ChemKin format \cite{chemkin}.  The number of accessible states can be very large even for a gas with a modest number of atoms.  To see this, we consider the initial state with $2m$ molecules of \ce{H2}, $m$ molecules of \ce{O2}, a single molecule of \ce{OH} (included to start the chain reaction process), and $20m$ atoms of \ce{Ar} at $1$~\si{atm} and $1{,}000$~\si{K}.  We refer to this as the $2m$\,\ce{H2} initial state and label it with $i=1$ throughout the article.  For this initial condition, we additionally impose a minimum temperature of $200$~\si{K} and a minimum pressure of $0.01$~\si{atm} for any state to be included.  Note that the $2m$\,\ce{H2} initial state is stoichiometrically balanced, so that all hydrogen in the mixture will be consumed during combustion.  The number of states accessible from the $6$\,\ce{H2} initial state (i.e., $m=3$) is just $N=347$, and the corresponding network has a modest complexity (Fig.~\ref{fig_scc3}, the first network at the top).  We note that this number is significantly smaller than the number of all possible states with the same numbers of \ce{H}, \ce{O}, and \ce{Ar} atoms ($842$ states, as determined by direct enumeration) due to the accessibility requirement and the thermodynamic constraints mentioned above.  The number of accessible states quickly grows as we scale up the number of molecules proportionally (Fig.~\ref{fig_scc3}), reaching $N=21{,}717$ for the $16$\,\ce{H2} initial state, with just $6m+2=50$ atoms (counting only \ce{O} and \ce{H}, since \ce{Ar} serves only as an energy buffer and does not actively participate in the reactions).  In fact, the number of states appears to grow with the number of atoms slightly faster than a power law with exponent $4.3$ (Fig.~\ref{dimension}(a), open circles) \cite{foot3}.

To compute the transition rates $W_{ij}$, we need the constants $\kappa_{nj}$ in Eq.~\eqref{wij}, which in this case is given by
\begin{equation}\label{kappa}
\kappa_{nj} = \frac{\kappa_n(T_j, p_j)}{\left(N_{\mathrm{A}} V\right)^{o_n-1}},
\end{equation}
where $o_n \equiv \sum_k \xi_{nk}$ is the molecularity, $N_{\mathrm{A}}$ is Avogadro's number, $V$ is the system volume, and $\kappa_n(T_j, p_j)$ are the rate constants defining the kinetic model (which for most reactions has a temperature dependence of the modified Arrhenius form, but may also involve efficiency coefficients for a third-body reaction and a pressure dependence of a falloff reaction).

\begin{figure*}[tp]
\includegraphics[width=0.7\textwidth]{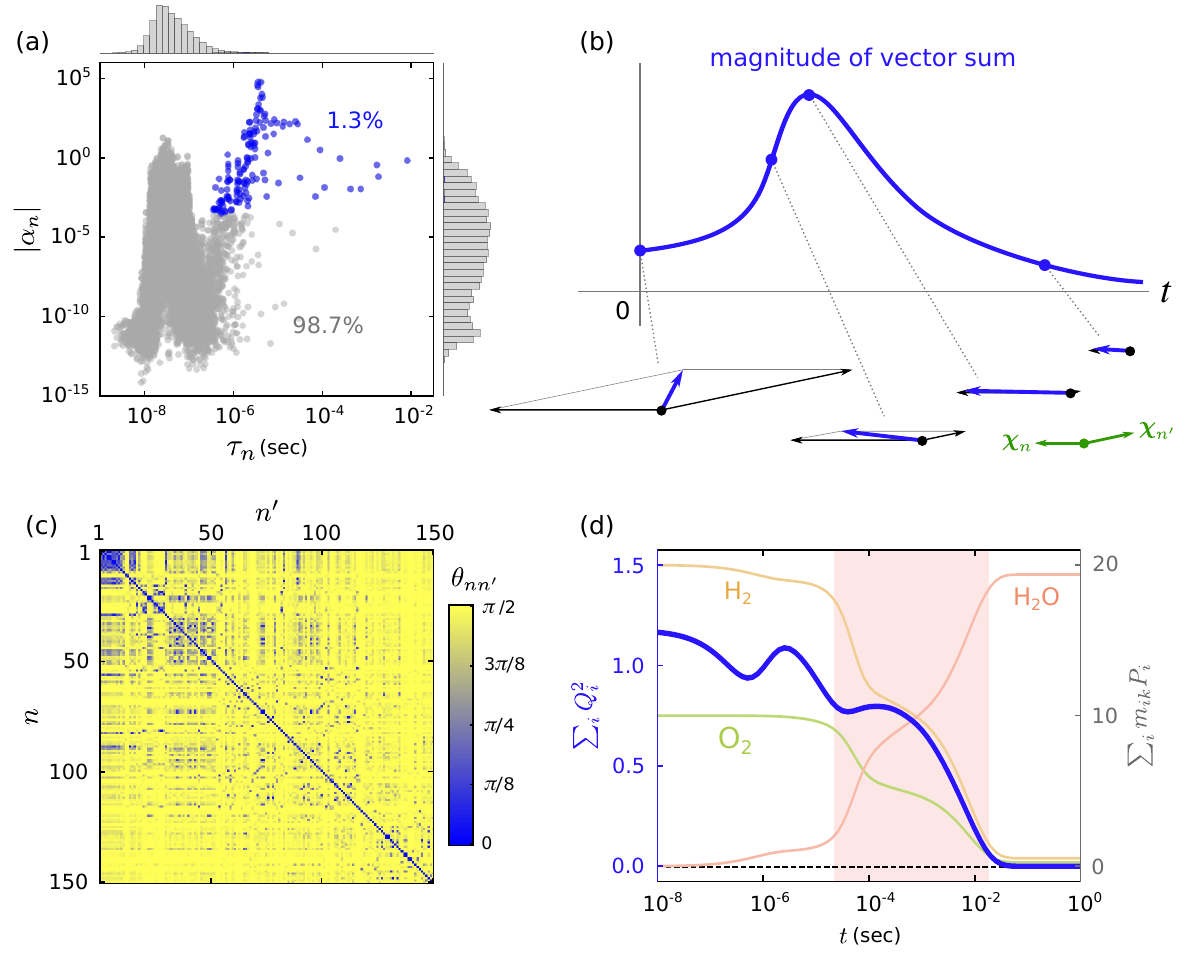}
\caption{Spectrum of the $20$\,\ce{H2} network in Fig.~\ref{fig1}(b).  (a)~Projection $\alpha_n$ onto eigenvectors vs.\ timescale $\tau_n$ associated with the corresponding eigenvalues.  Highlighted in blue are the $150$ eigenvalues with largest $\alpha_n$ among the $1{,}000$ eigenvalues with the largest $\tau_n$.  Histograms for $\alpha_n$ and $\tau_n$ are shown on the right and the top of the panel, respectively.  In each case, the histogram is shown separately for the blue and gray eigenvalues.  (b)~Illustration of non-monotonic dynamics due to a pair of nearly parallel eigenvectors $\boldsymbol{\chi}_n$ and $\boldsymbol{\chi}_{n'}$ (green arrows).  When the eigenvectors have large coefficients ($\alpha_n, \alpha_{n'} \gg 1$) and associated with very different timescales ($\tau_n \gg\tau_{n'}$), the sum of the corresponding terms in Eq.~\eqref{eqn:Q} (blue arrows) initially grows but eventually shrinks to zero.  (c)~Angle $\theta_{nn'}$ between the $n$th and $n'$th eigenvectors among those corresponding to the blue dots in (a), sorted by descending magnitude of their projection coefficients $\alpha_n$ and $\alpha_{n'}$.  (d)~Non-monotonic decay of $\sum_i Q_i^2$ for the network in Fig.~\ref{fig1}(b).  For reference, we also plot the expected number of molecules of reactants (\ce{H2} and \ce{O2}) and product (\ce{H2O}) as functions of time, showing the progression of the combustion process. The red shading indicates the interval of the ignition event (defined in the text).
\label{fig2}}
\end{figure*}

Since the transition rates $W_{ij}$ span multiple orders of magnitude, the flow of probability under Eq.~\eqref{cme} is typically limited to a small portion of the network.  This observation can be used upfront to reduce the size of the network, and thus the computational burden, without significantly affecting the dynamics.  For this purpose, we apply thresholding in our state enumeration algorithm; if there is a transition from state $j$ to a new state $i$, then state $i$ is added only if the rate of that transition is not negligible compared to the rate of the opposite transition, i.e., if $W_{ij} > \epsilon W_{ji}$, where $\epsilon$ is a (small) threshold~\cite{foot2}.  Thus, if there are multiple states $j$ from which the system transitions to state $i$, we add state $i$ if and only if at least one of these transitions carries non-negligible flow of probability into state $i$.  Once a full list of states is obtained, both $W_{ij}$ and $W_{ji}$ are computed for each pair of enumerated states $i$ and $j$.  Unless otherwise noted, we employ $\epsilon=10^{-3}$ in the remainder of the paper, which is small enough for accurate computation of system trajectories (see reference comment~\cite{error_comment} for details).  For example, for the $20$\,\ce{H2} initial state, this $\epsilon$-thresholding reduces the size of the network by more than $82$\% ($N = 63{,}835 \to 11{,}129$).  This reduced $20$\,\ce{H2} network and the corresponding $11{,}106$-dimensional transition rate matrix $\mathbf{W}\equiv(W_{ij})_{1 \le i,j \le N}$ are visualized in Fig.~\ref{fig1}(b) and (c), respectively.  The $\epsilon$-thresholding consistently yields significant reduction in $N$ and in the complexity of the network, as seen by comparing the open and filled circles in Fig.~\ref{dimension}(a).  Furthermore, the percentage reduction in $N$ achieved by this procedure grows with the number of atoms in the system, as shown in Fig.~\ref{dimension}(b).  With this reduction technique, we were able to consider network sizes as large as $N=74{,}421$, corresponding to the $30$\,\ce{H2} initial state with $92$ atoms (counting \ce{O} and \ce{H}).

\section{Spectral analysis}

The linearity of Eq.~\eqref{cme} implies that all information about the dynamics is encoded in the eigenvalues and eigenvectors of the matrix $\mathbf{W}$.  We first note that the structure of the matrix guarantees the sum over each column to be zero, i.e., $\sum_i W_{ij}=0$ for all $j$.  Using this and applying the Gershgorin Circle Theorem~\cite{Horn:1990}, it follows that zero is an eigenvalue of $\mathbf{W}$ (which can be degenerate), and that all the other eigenvalues have strictly negative real parts.  Moreover, it can be shown that each right eigenvector associated with the zero eigenvalue can be normalized so that its components are all non-negative and sum to unity (see Appendix).  Each such normalized eigenvector, or a convex combination of multiple such vectors, is thus a steady-state probability distribution for Eq.~\eqref{cme}.

If the network of states is strongly connected, or equivalently if the matrix $\mathbf{W}$ is irreducible, the Perron-Frobenius Theorem~\cite{Horn:1990} implies that the zero eigenvalue is actually non-degenerate and that the components of its right eigenvector are all strictly positive, leading to a unique steady-state distribution with $P_i>0$ for all $i$.  This holds true for the \ce{H2}/\ce{O2} combustion process considered here, since all reactions in that process are reversible, making the networks of states undirected and thus strongly connected.  Note, however, that many of the reactions have tiny reverse reaction rate, which leads to many states with low probability $P_i$, as we will show below.

The temporal evolution of the system under Eq.~\eqref{cme} can be decomposed into eigenmodes.  Assuming that the network is strongly connected, we denote by $\chi_{0i}$ the normalized right eigenvector associated with the zero eigenvalue, i.e., the unique steady-state distribution.  For the remaining $N-1$ eigenvalues, we use $\lambda_n$ to denote the $n$th eigenvalue (in an arbitrary order) and $\chi_{ni}$ to denote the $i$th component of the corresponding right eigenvector normalized to unit length in $2$-norm.  Given the initial probability distribution $P_i(0)$, the deviation from the steady-state distribution, $Q_i(t)\equiv P_i(t) - \chi_{0i}$, can be expressed in terms of
the (generally non-orthogonal) projection onto the eigenbasis,
\begin{equation}\label{eqn:Q}
Q_i(t) = \sum_{n>0} \alpha_n {\mathrm e}^{\lambda_n t} \chi_{ni}, \text{ where } \alpha_n \equiv \sum_i \eta_{ni}P_i(0),
\end{equation}
and $\eta_{ni}$ is the $i$th component of the left eigenvector associated with $\lambda_n$. Here, the left eigenvectors are normalized so as to satisfy $\sum_i \eta_{ni} \chi_{n'i} = \delta_{nn'}$, where we recall that $\delta_{nn'}$ denotes the Kronecker delta.  In this eigen-decomposition, all terms decay monotonically in time because $\lambda_n < 0$ for all $n>0$, implying that $P_i(t)\to \chi_{0i}$ as $t\to\infty$, i.e., the system converges to the steady-state distribution.  The contributions of these eigenmodes are quantified by the mode strength coefficients $\alpha_n$, with their decay characterized by the timescale parameters $\tau_n \equiv -1/\text{Re}(\lambda_n) > 0$.

The multi-scale nature of the \ce{H2}/\ce{O2} combustion process is apparent in the distributions of $\alpha_n$ and $\tau_n$ shown in Fig.~\ref{fig2}(a), spanning many orders of magnitude.  On the one hand, the vast majority of the eigenmodes decay very fast as $\tau_n$ is tiny and thus contribute very little to the overall dynamics.  Indeed, the probability $P_i(t)$ computed using only the $150$ modes highlighted in blue in Fig.~\ref{fig2}(a) stays within $3\times 10^{-4}$ of that computed using all eigenmodes for all $i$ and all $t$, while  $P_i(t)$ computed using only the $1{,}000$ eigenmodes with largest $\tau_n$ stays within $2\times10^{-6}$ of that computed using all eigenmodes (code for efficient trajectory computation available in our online repository~\cite{github}).  On the other hand, there are several modes with slow timescales and significant $\alpha_n$.  These features are shared with sloppy models~\cite{2015_Transtrum,2016_White}, which have modes with strength and timescale varying over many orders of magnitude but can describe the process accurately with only a handful of these modes. 

However, if the matrix $\mathbf{W}$ is \textit{non-normal} (i.e., if the matrix normality condition $\sum_k W_{ik}W_{jk} = \sum_k W_{ki}W_{kj}, \forall i,j$ is violated), the monotonic decay of individual eigenmodes does not tell the whole story.  For a normal $\mathbf{W}$, the right eigenvectors are all orthogonal to each other.  It then follows from the eigen-decomposition in Eq.~\eqref{eqn:Q} that $Q_i(t)$ decays monotonically in $2$-norm, i.e., $\sum_i Q_i^2 \to 0$ as $t\to\infty$.  For a non-normal $\mathbf{W}$, the eigenvectors are not necessarily orthogonal to each other and can be nearly parallel (or even completely parallel, which is equivalent to having a degenerate eigenvector).  Consider a system whose $n$th and $n'$th eigenvectors are nearly parallel, and suppose that the coefficients $\alpha_n$ and $\alpha_{n'}$ have opposite signs with large magnitudes, and that the timescales $\tau_n$ and $\tau_{n'}$ are very different.  Then, the sum of the corresponding terms in Eq.~\eqref{eqn:Q} will exhibit highly non-monotonic dynamics, initially growing to a large size before eventually shrinking to zero, as illustrated in Fig.~\ref{fig2}(b).  For the \ce{H2}/\ce{O2} combustion network of Fig.~\ref{fig1}, the matrix $\mathbf{W}$ is highly non-normal, with many eigenvectors having small angles between them (Fig.~\ref{fig2}(c)), and we indeed observe $\sum_i Q_i^2$ to behave non-monotonically (Fig.~\ref{fig2}(d), blue curve).  The non-monotonic dynamics emerge before the ignition event, which starts at $t \approx 2.2 \times 10^{-5}$~sec and ends at $t \approx 1.9 \times 10^{-2}$~sec (Fig.~\ref{fig2}(d), red shading).  Here, the ignition event is defined as the interval in which between $5$\% and $95$\% of the total temperature change is observed.
The non-monotonicity continues during the conversion of \ce{H2} and \ce{O2} to \ce{H2O}, until about half of the fuel is consumed at $t \approx 10^{-3}$~sec.

\section{Non-monotonic entropy dynamics}

The non-monotonic decay described above for non-normal $\mathbf{W}$ is closely related to the so-called R\'{e}nyi entropy $H_2\equiv -\log(\sum_i P_i^2)$~\cite{renyi_1961}.  Like the Shannon entropy $H_1\equiv -\sum_i P_i\log(P_i)$, the R\'{e}nyi entropy quantifies the spread of the probability distribution and can thus be regarded as a measure of the uncertainty about the state of the system.  Just as Boltzmann's H-theorem guarantees monotonic evolution of the Shannon entropy when a molecular chaos assumption is satisfied (which would entail $W_{ij}=W_{ji}$ in this setting)~\cite{Tolman:1979}, the R\'{e}nyi entropy is monotonic when the normality condition $\sum_k W_{ik}W_{jk} = \sum_k W_{ki}W_{kj}$ is satisfied, since the orthogonality of the eigenvectors guarantees monotonic decay \cite{foot1}.  To illustrate this, we first note that the following formula can be derived:
\begin{equation}\label{eqn:Renyi_enropy}
\frac{{\mathrm d}H_2}{{\mathrm d}t} = -2\cdot\frac{\sum_{ij} P_i S_{ij} P_j}{\sum_i P_i^2},
\end{equation}
where $\mathbf{S} \equiv (S_{ij})_{1 \le i,j \le N}$, $S_{ij} \equiv \frac{1}{2}(W_{ij}+W_{ji})$ is the symmetric part of $\mathbf{W}$.  This equation establishes a connection between information theory and dynamical systems theory: the rate of change of the R\'{e}nyi entropy is directly proportional to the Rayleigh quotient of $\mathbf{S}$, which is of interest in non-normal growth analysis~\cite{Neubert:1997,2005_Trefethen}.  This connection reflects the fact that the R\'{e}nyi entropy is a function of the 2-norm of the probability distribution $P_i$.  In particular, Eq.~\eqref{eqn:Renyi_enropy} implies that the minimum possible $\mathrm{d}H_2/\mathrm{d}t$ is determined by the largest eigenvalue of $\mathbf{S}$, which is non-positive and known as the \textit{reactivity} (up to the factor of $-2$).  If $\mathbf{W}$ is normal, it is known that the reactivity equals the largest eigenvalue of $\mathbf{W}$, which is zero.  This implies that $\mathrm{d}H_2/\mathrm{d}t \ge 0$, i.e., the R\'{e}nyi entropy can only increase during the system's evolution towards the steady-state distribution.  If $\mathbf{W}$ is non-normal, the reactivity can be strictly positive, meaning that the R\'{e}nyi entropy decreases.  The leading eigenvector of $\mathbf{S}$ gives the distribution maximizing the rate of decrease.

\begin{figure}[t]
\includegraphics[width=\columnwidth]{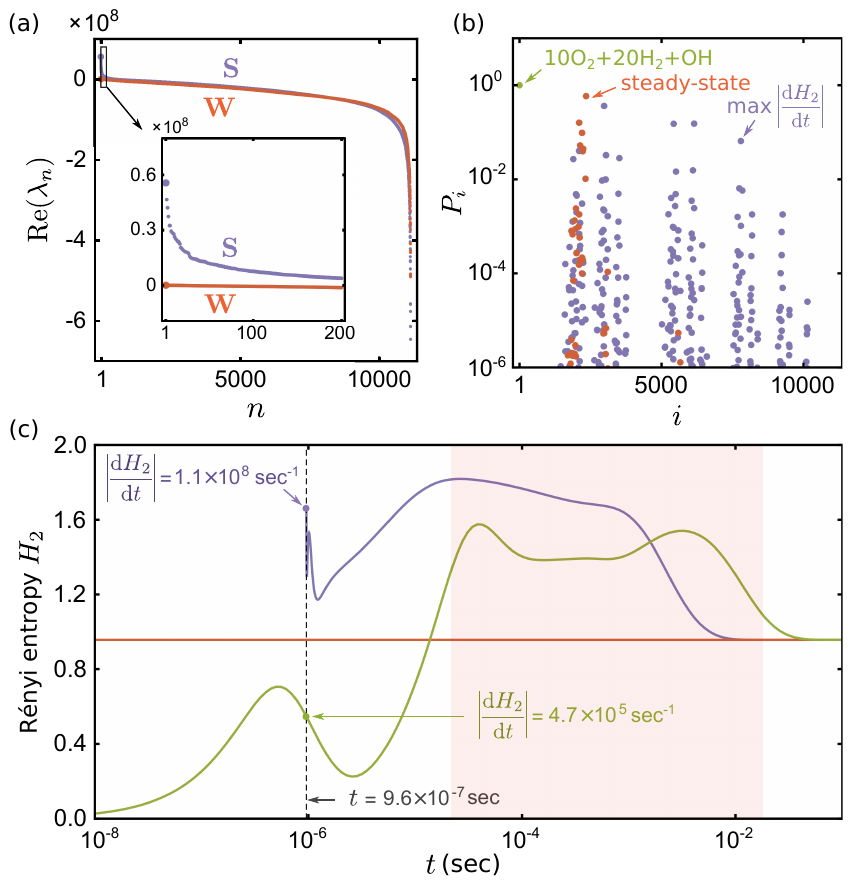}
\vspace{-7mm}
\caption{Non-monotonic entropy dynamics for the $20$\,\ce{H2} network of Fig.~\ref{fig1}(b).  (a)~Spectra of the non-normal matrix $\mathbf{W}$ and its symmetric part $\mathbf{S}$.  While the leading eigenvalue of $\mathbf{W}$ is zero (larger orange dot), the leading eigenvalue of $\mathbf{S}$ is strictly positive (larger purple dot), indicating that the R\'{e}nyi entropy $H_2$ can decrease, with the fastest rate $\lvert\mathrm{d}H_2/\mathrm{d}t\rvert$ determined by that eigenvalue.  (b)~Steady-state distribution (orange) and the distribution with the maximum $\lvert\mathrm{d}H_2/\mathrm{d}t\rvert$ (purple), constructed by normalizing the leading eigenvectors of $\mathbf{W}$ and $\mathbf{S}$, respectively.  Also shown is the single-state distribution corresponding to the initial condition used to construct the network (green).  (c)~Evolution of $H_2$ starting from the three distributions in (b).  The green trajectory achieves its maximum $\lvert\mathrm{d}H_2/\mathrm{d}t\rvert$ at $t=9.6\times10^{-7}$~sec (green dot), while the purple trajectory (shifted forward in time to facilitate comparison) initially exhibit two orders of magnitude faster decrease in $H_2$ (purple dot).  The red shading indicates the same interval of ignition event shown in Fig.~\ref{fig2}(d).
\label{fig_entropy}}
\end{figure}

For the \ce{H2}/\ce{O2} combustion network of Fig.~\ref{fig1}, the reactivity is indeed strictly positive (Fig.~\ref{fig_entropy}(a), larger purple dot), indicating the existence of a probability distribution for which the system exhibits the maximum possible $\mathrm{d}H_2/\mathrm{d}t$ (Fig.~\ref{fig_entropy}(b), smaller purple dots).  Such a distribution can be constructed by properly normalizing the corresponding eigenvector of $\mathbf{S}$, which is guaranteed to be possible by the Peron-Frobenius theorem.  The system evolution starting from that distribution indeed exhibits the predicted rate of entropy decrease, $\mathrm{d}H_2/\mathrm{d}t=-1.1\times10^8$~sec$^{-1}$ (Fig.~\ref{fig_entropy}(c), purple dot), reflecting the quick focusing of the probability initially spread over many states onto a small number of states.  This distribution, however, is not the only one with $\mathrm{d}H_2/\mathrm{d}t<0$, as there are many other strictly positive eigenvalues for $\mathbf{S}$ (Fig.~\ref{fig_entropy}(a)), from which distributions sharing the same property can be constructed.  The decreasing $H_2$ (albeit at a much slower rate than the maximum) is also observed during the system evolution starting from the $20$\,\ce{H2} initial state, with several periods of $\mathrm{d}H_2/\mathrm{d}t<0$ over the non-monotonic trajectory (Fig.~\ref{fig_entropy}(c), green curve).  In particular, the fastest decrease occurs at $t=9.6\times10^{-7}$~sec, before the ignition event, indicated by the red shading in Fig.~\ref{fig_entropy}(c).  This suggests the existence of bottlenecks in the network of states, where non-normality directs the flow of probability to accumulate on a small number of states during the pre-ignition dynamics and sets up the stage for the explosive ignition dynamics.

\section{Condition for non-monotonic species dynamics}
\label{sec:non-monotonicity-condtion}

\begin{figure}[t]
\includegraphics[width=\columnwidth]{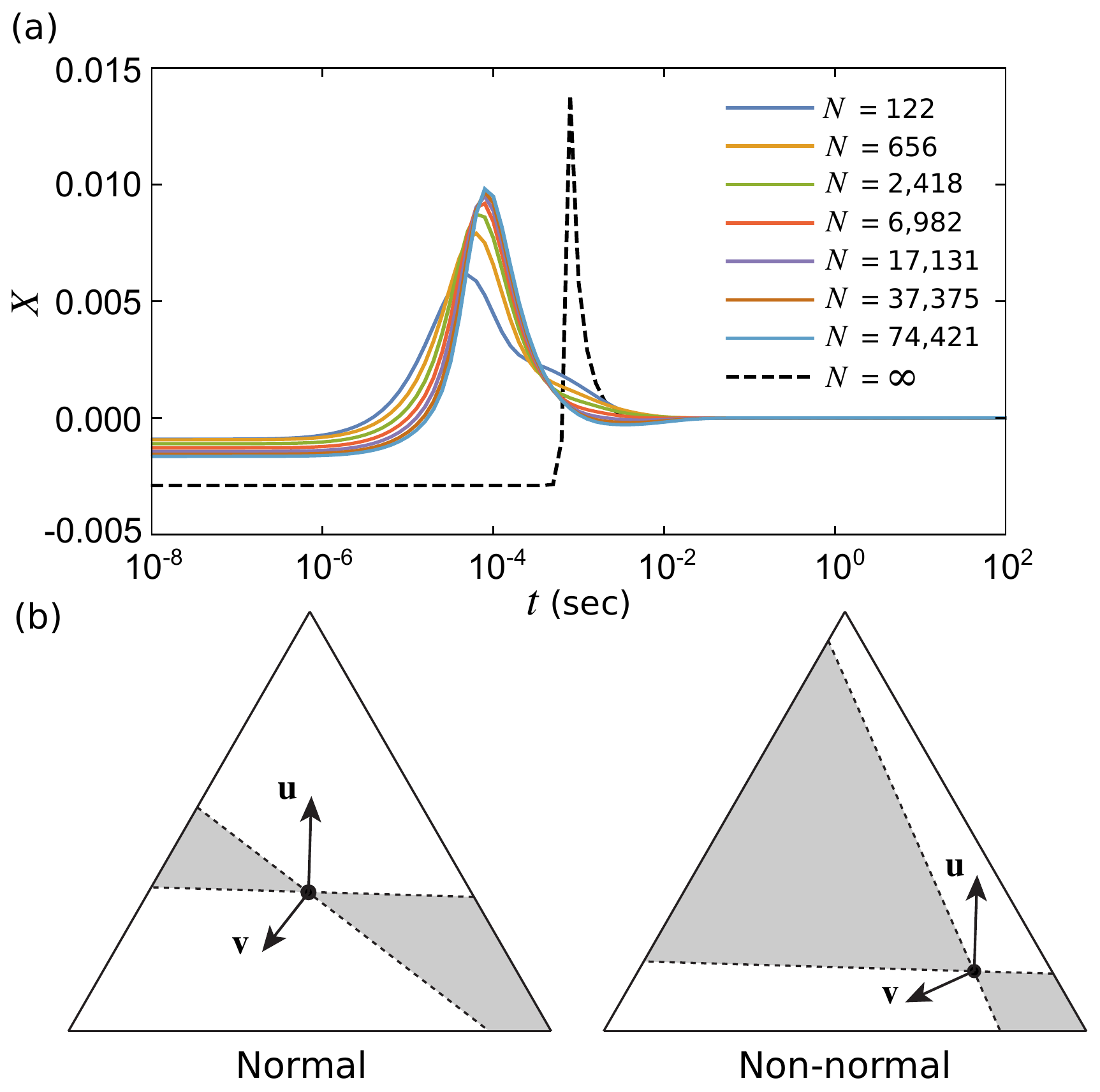}
\caption{Non-monotonic chemical species dynamics.  (a)~Expected fraction of molecules that are radicals in the \ce{H2}/\ce{O2} combustion process, as a function of time for several different network sizes $N$.  The seven solid curves correspond to the $6$\,\ce{H2}, $10$\,\ce{H2}, $14$\,\ce{H2}, \ldots, $30$\,\ce{H2} networks, whose sizes $N$ are shown in the plot.  The dashed curve represents the continuum limit $N = \infty$.  (b)~$(N-1)$-simplex representing the space of all possible probability distributions $P_i$ for $N=3$.  The dashed lines, orthogonal to the vectors $\mathbf{u}$ and $\mathbf{v}$, divide the simplex into quadrants, with their intersection corresponding to the steady-state distribution $\chi_{0i}$ (which would be at the center of the triangle if $\mathbf{W}$ is normal.  Trajectories starting in a gray quadrant exhibits non-monotonic dynamics. 
\label{fig3}}
\end{figure}

We now derive an analytical, geometrically interpretable condition for individual species to exhibit non-monotonic dynamics.  Given a subset of species $\mathcal{X}$, let $X=X(t)$ denote the expected fraction of molecules that are of the species in $\mathcal{X}$ at time $t$, relative to its value for the steady-state distribution.  This quantity can be expressed as $X(t) = \sum_i  M_i Q_i$, where $M_i \equiv \sum_{k\in\mathcal{X}}m_{ik}/\sum_k m_{ik}$ is the normalized counts of molecules of the species in $\mathcal{X}$ for state $i$.  Using this formula, the time evolution of the species can be calculated from the solutions of Eq.~\eqref{cme}.  For example, the evolution of the radicals for the \ce{H2}/\ce{O2} combustion process computed this way is shown for different initial number of molecules (and thus different $N$) in Fig.~\ref{fig3}(a).  As $N$ increases, the trajectory appears to approach the continuum limit determined by Cantera.  Thus, our results suggest that the sharp peaks of radical concentrations associated with ignition observed in the continuum limit have origin in non-monotonic dynamics promoted by the non-normality of the transition rate matrix $\mathbf{W}$ in the chemical master equation. 

Since the probabilities $P_i$ are all strictly positive and sum to unity, the trajectory of the corresponding point under Eq.~\eqref{cme} is limited to the set $\{ (P_1,\ldots,P_N) \,\vert\, P_i \ge 0, \forall i \text{ and } \sum_i P_i = 1 \}$, which is an $(N-1)$-simplex (a generalization of a triangle).  This simplex is illustrated in Fig.~\ref{fig3}(b) for $N=3$, in which case it is a triangle.  Each vertex of the simplex corresponds to the distribution with all the probability concentrated on a single state.  Each face (or edge for $N=3$) corresponds to distributions with the probability spread over multiple states.  The interior corresponds to distributions with non-zero probability for all states.  Trajectories determined by Eq.~\eqref{cme} travel within this simplex and eventually approach the point corresponding to the steady-state distribution $\chi_{0i}$.  If the matrix $\mathbf{W}$ is normal, this point would be at the center of the simplex, since the vector of all ones is a right eigenvector of $\mathbf{W}$ (in addition to being a left eigenvector), which implies that the steady-state distribution is uniform.  However, if $\mathbf{W}$ is non-normal, the point could be anywhere in the simplex and can be close to the boundaries of the simplex, since $\chi_{0i}$ can be non-uniform.  Indeed, for the \ce{H2}/\ce{O2} combustion network of Fig.~\ref{fig1}, it lies very close to a $24$-dimensional hyper-edge, with a $2$-norm distance $\approx 1.4\times10^{-5}$ ($\approx 0.01$\% of the distance from the center of the $11{,}129$-dimensional simplex to the hyper-edge).  This reflects the property of the steady-state distribution that it is highly localized: more than $99.99$\% of the probability is concentrated on the corresponding $24$ states (mostly composed of \ce{H2O}, with just a few molecules of \ce{H2}, \ce{O2}, and other radicals).

\begin{figure*}[t]
\includegraphics[width=\textwidth]{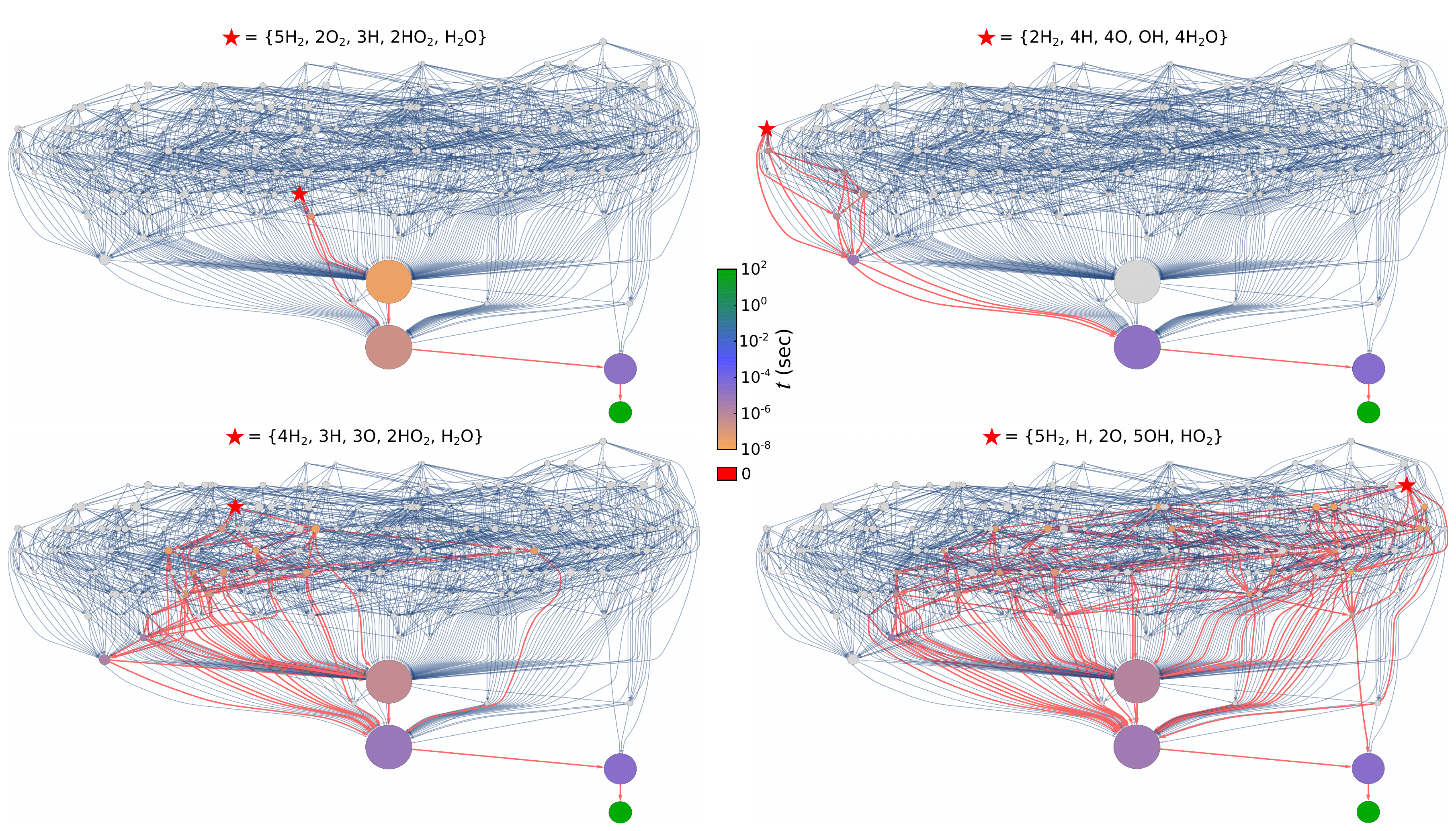}
\vspace{-7mm}
\caption{Flow of probability over the DAG structure in the \ce{H2}/\ce{O2} combustion network constructed from the $8$\,\ce{H2} initial state.  In each panel, the network is visualized as in Fig.~\ref{fig_scc3}, except that the initial probability distribution is fully concentrated on a randomly chosen state (red star symbol).  During system evolution, probability branches out and flows downward along various paths (red arrows; showing only those with probabilities $<10^{-3}$), eventually converging to the SCC at the bottom (green circle), which supports the steady-state distribution.  Note that the four randomly chosen initial states are different from the $8$\,\ce{H2} initial state used to construct the network.
\label{fig_rand_ic}}
\end{figure*}

To derive the condition for non-monotonic dynamics, we first note that $X$ converges to zero as $t\to\infty$, since $X$ is defined relative to the steady-state distribution.  Thus, the condition for $X$ to initially move away from zero before converging to zero is that $X$ and its derivative ${\mathrm d}X/{\mathrm d}t$, given by ${\mathrm d}X/{\mathrm d}t=\sum_{ij}  M_i W_{ij} Q_j$, have the same sign.  To rewrite this condition in a geometrically interpretable form, we define vectors $\mathbf{u}$ and $\mathbf{v}$ as those with the $i$th component $u_i \equiv \sum_j M_j O_{ji} $ and $v_i \equiv \sum_{jk} M_j W_{jk}O_{ki}$, respectively, where $O_{ij} \equiv \delta_{ij}-1/N$.  The vector $\mathbf{u}$ can be interpreted as the projection of the vector $\mathbf{m} \equiv (M_i)$ onto the simplex.  Likewise, $\mathbf{v}$ is the projection of the vector $\mathbf{m}' \equiv (M'_k)$, $M'_k \equiv \sum_j M_j W_{jk}$, onto the simplex.  Using these vectors, we can rewrite $X$ and ${\mathrm d}X/{\mathrm d}t$ as $X = \sum_i u_i Q_i$ and ${\mathrm d}X/{\mathrm d}t = \sum_i v_i Q_i$.  The simplex can be divided into ``quadrants'' by two hyperplanes: one orthogonal to $\mathbf{u}$ (given by $\sum_i u_i Q_i=0$) and the other orthogonal to $\mathbf{v}$ (given by $\sum_i v_i Q_i=0$).  Both hyperplanes contain the point $Q_i=0$, which corresponds to the steady-state distribution $\chi_{0i}$.  Then, a geometric condition for exhibiting non-monotonic dynamics after time $t$ is that the point $Q_i(t)$ lies in either the quadrant $\bigl\{ \sum_i u_i Q_i > 0 \text{ and } \sum_i v_i Q_i > 0 \bigr\}$ or the quadrant $\bigl\{ \sum_i u_i Q_i < 0 \text{ and } \sum_i v_i Q_i < 0 \bigr\}$.  These quadrants are shaded in gray in Fig.~\ref{fig3}(b).  The vectors $\mathbf{u}$ and $\mathbf{v}$, as well as the quadrants of non-monotonicity, can also be defined and are shown in Fig.~\ref{fig_non_monotonic} for normal and non-normal two-dimensional systems.  (We note that $\mathbf{u}$ and $\mathbf{v}$ for these systems do not require the projection by $O_{ij}$, since their states are not constrained to a simplex.)

The condition just derived shows that, for both normal and non-normal $\mathbf{W}$, there are initial distributions leading to non-monotonic behavior of $X$.  This means that, even though the $2$-norm $\sum_i Q_i^2$ converges to zero monotonically for normal $\mathbf{W}$ (as mentioned earlier), the convergence of the projection $X = \sum_i u_i Q_i$ may be non-monotonic.  We thus conclude that non-normality of $\mathbf{W}$ is not a requirement for non-monotonicity of $X$.  However, non-normal $\mathbf{W}$ is different from normal $\mathbf{W}$ in that the steady-state distribution tends to lie near the boundaries of the simplex.  This implies that the quadrants of non-monotonicity occupy larger fraction of the simplex than for normal $\mathbf{W}$, indicating that non-normal $\mathbf{W}$ is more likely to exhibit non-monotonic dynamics.  Moreover, the non-monotonicity tends to be more pronounced for non-normal $\mathbf{W}$, since the further away the initial point is from the hyperplane orthogonal to $\mathbf{v}$, the larger is the rate at which $X$ moves away from zero (recalling that ${\mathrm d}X/{\mathrm d}t = \sum_i v_i Q_i$ is the distance from the hyperplane).  For our \ce{H2}/\ce{O2} combustion example, we indeed observe a sharp transient increase in the expected number of radical molecules, as we saw in Fig.~\ref{fig3}(a).

\section{Connected component analysis}
\label{sec:cca}

In general, the network of states representing the system~\eqref{cme} is directed and can be decomposed into strongly connected components (SCC), where an SCC is defined as a subset of nodes in which a directed path exists between every node pair in both directions.  This allows for a coarse-grained representation of the network by a (different) network of SCCs.  In this representation, which we refer to as the \textit{SCC network}, each SCC is regarded as a node, and a directed link is drawn from one SCC to another if, in the original network, there is a directed link from some node in the first SCC to a node in the second SCC.  It follows from a well-known fact from graph theory that the SCC network must be a DAG, i.e., it does not contain any closed loop.

For the \ce{H2}/\ce{O2} combustion, the whole network is actually strongly connected (since all the reactions are modeled as a reversible one) and thus has just one SCC.  However, a non-trivial DAG structure emerges when we take into account the link weights $W_{ij}$, many of which are negligibly small compared to the weights $W_{ji}$ of the reciprocal links.  This is because the reaction giving rise to a link is often nearly irreversible under the given condition in the sense that its rate in one direction is orders of magnitude larger than in the other direction.  To capture this link directionality, we remove all transitions from state $j$ to state $i$ satisfying $W_{ij} < \mu W_{ji}$ for a given small constant $\mu>0$.  We note that this $\mu$-thresholding prunes links representing negligibly rare transitions, while the $\epsilon$-thresholding introduced above prunes nodes representing rarely visited states (together with the links involving those nodes).  Once the $\mu$-thresholding is applied, we decompose the resulting network into SCCs, revealing a DAG structure.  Throughout this article, we use $\mu=2\times10^{-3}$.  With this value, we find that the $\mu$-thresholding removes much fewer links than the $\epsilon$-thresholding ($4.4$\% compared to $86.3$\% of all links in the $20$\,\ce{H2} network constructed with no $\epsilon$- or $\mu$-thresholding) and does not significantly affect the dynamics (with maximum error in $P_i(t)$ less than $5\times10^{-3}$ for all $i$ and all $t$).  Because the SCC network has a DAG structure, one can arrange the SCCs in horizontal layers in such a way that the directions of all links, and thus flows of probability under Eq.~\eqref{cme}, are downwards.  This layered SCC arrangement is used in Fig.~\ref{fig_scc3} for the $6$\,\ce{H2}, $8$\,\ce{H2}, \ldots, $16$\,\ce{H2} networks constructed without the $\epsilon$-thresholding but with $\mu$-thresholding.  The DAG structure dictates the downward flow of probability from an arbitrary initial distribution to the final steady-state distribution contained in the bottom SCC, as shown in Fig.~\ref{fig_rand_ic} for randomly chosen initial state in the $8$\,\ce{H2} network.  Code for computing the SCCs, as well as data files describing individual states, is available in our online repository~\cite{github}.

\begin{figure}
\includegraphics[width=\columnwidth]{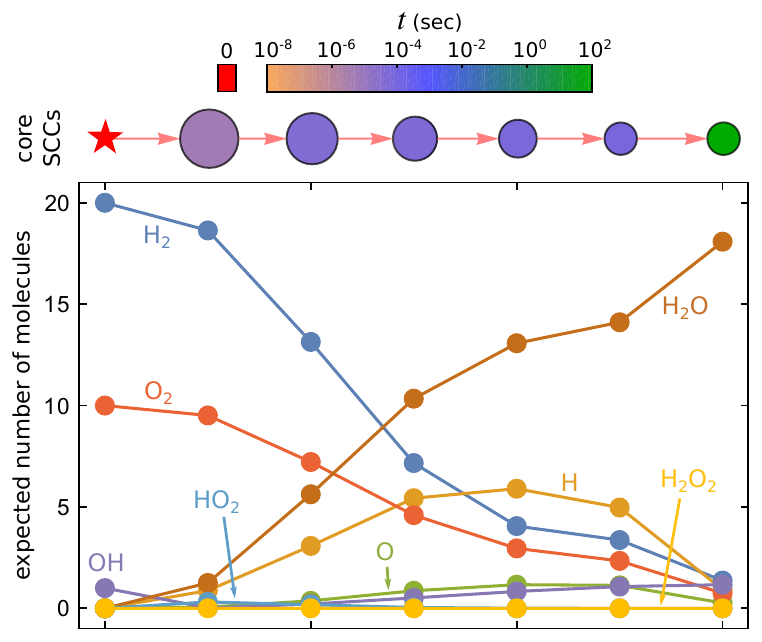}
\caption{Molecular composition profiles in the linear chain of core SCCs in the $20$\,\ce{H2} network in Fig.~\ref{fig1}(b).  For a given SCC, we show the expected number of each chemical species in a state belonging to that SCC at the time of maximum probability (encoded as the node color).  We observe that, while the numbers of reactant molecules (\ce{H2} and \ce{O2}) monotonically decrease and the number of product molecules (\ce{H2O}) monotonically increases along the linear chain, some of the radicals (\ce{H} and \ce{O}) have peaks for intermediate SCCs.
\label{fig_scc}}
\end{figure}

\begin{figure*}[t]
\includegraphics[width=2\columnwidth]{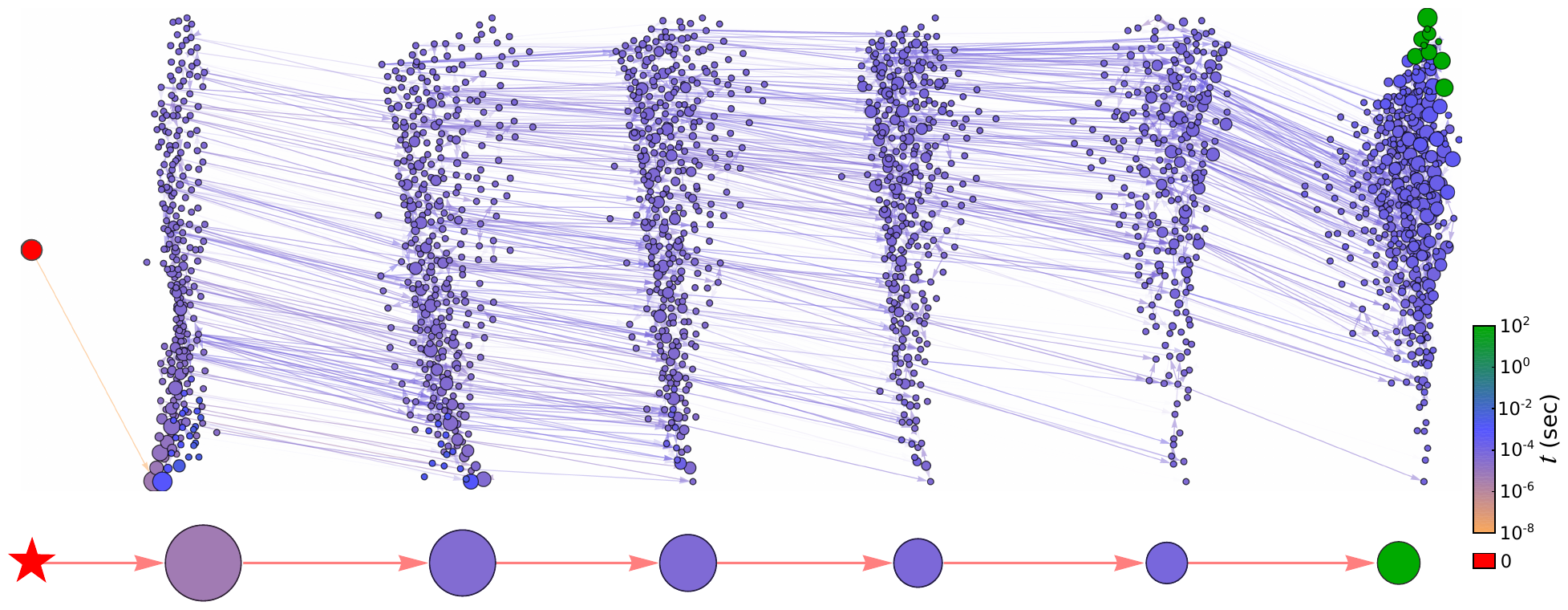}
\caption{Detailed connectivity structure within and between the core SCCs in the linear chain from Fig.~\ref{fig_scc} (reproduced here at the bottom).  The size of each node $i$ is proportional to the maximum of $P_i(t)$ over $t$, and the color coding for the nodes is the same as in Fig.~\ref{fig_scc}.  Nodes with negligibly small probability ($P_i(t)<10^{-8}$ for all $t$) are excluded to avoid overcrowding. 
\label{fig_scc2}}
\end{figure*}

We observe that the initial condition used to construct the network tends to belong to an SCC located toward the bottom of the layered DAG structure (the red stars in Fig.~\ref{fig_scc3}).  If we focus now on the SCCs that are accessible from that initial state, we find that these \textit{core} SCCs form a linear chain leading to the SCC containing the steady-state distribution (color-coded circles in Fig.~\ref{fig_scc3}, excluding SCCs with $P_i < 10^{-3}$ for all $t$).  The same linear chain is obtained also if we take the network constructed with the $\epsilon$-thresholding and then remove insignificant reverse transitions using the $\mu$-thresholding.  While the expected number of the reactants \ce{H2} and \ce{O2} decay monotonically and the product \ce{H2O} increase monotonically along the chain, the numbers of radical molecules change non-monotonically, as shown in Fig.~\ref{fig_scc} for the $20$\,\ce{H2} network.  While this coarse-grained view of the network summarizes the dynamics with a simple linear chain, complex network structures exist both within and between the individual SCCs in the chain, as shown in Fig.~\ref{fig_scc2}.  This full network is the same as that in Fig.~\ref{fig1}(b), except that it is visualized using the linear chain structure in Fig.~\ref{fig_scc}.

The DAG structure of the SCC network has a direct implication for the steady-state distribution $\chi_{0i}$.  Indeed, we can show that $P_i=\chi_{0i}$ is strictly positive only for nodes $i$ belonging to the SCCs without any outgoing links, which can always be placed at the bottom of the layered arrangement (see Appendix for a rigorous proof).  Since the number and the sizes of such SCCs are often small, this implies that the steady-state distribution is localized.  For the networks in Fig.~\ref{fig_scc3}, the distribution is highly localized, with all probability concentrated in the green SCC at the bottom of each network (and further localized within that SCC, as illustrated in Fig.~\ref{fig_scc2} for the $20$\,\ce{H2} network).  Moreover, we observe a similar localization even when the $\mu$-thresholding is not applied and the whole network forms a single SCC, as we saw in the previous section, reflecting the fact that the dynamics are essentially unaltered by the thresholding.

\section{Random networks}

\begin{figure*}[t]
\includegraphics[width=\textwidth]{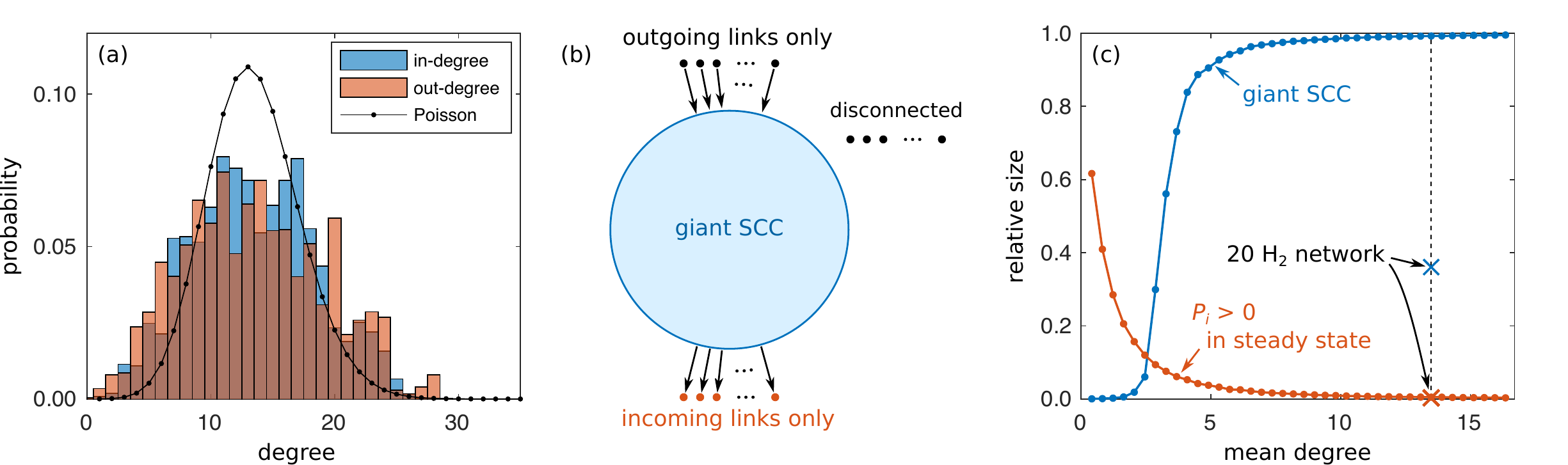}
\caption{Topological comparison between the $20$\,\ce{H2} network in Fig.~\ref{fig1}(b) and various classes of random networks.  (a)~In- and out-degree distributions of the combustion network (blue and orange histograms, respectively), compared to the Poisson distribution (black curve).  (b)~Typical SCC structure of a random chemical reaction network with topological parameters matching those of the combustion network.  (c)~Relative size of the giant SCC (blue) and relative number of nodes with $P_i>0$ in steady-state distribution (orange) for random chemical reaction networks, as a function of the mean degree.  The values of these quantities for the combustion networks are indicated by cross symbols.  The curves are averaged over $100$ network realizations.
\label{fig_percolation}}
\end{figure*}

To interpret the linear chain structure shown in Figs.~\ref{fig_scc} and \ref{fig_scc2} for the ($\mu$-thresholded) $20$\,\ce{H2} network, we now compare its topological properties to those of generic directed random networks.  We first consider fully random networks with the same numbers of nodes $N=11{,}129$ and the same number of directed links $M=150{,}077$.  Such a random network consists of a single giant SCC containing all $N$ nodes with very high probability ($96.8$\%, estimated from $1{,}000$ realizations), and the average size of the giant SCC is $\approx 11{,}128.97 \pm 0.18$.  The giant SCC appears because the average degree $d \equiv M/N \approx 13.5$ is far above the percolation threshold of $d=1$~\cite{Dorogovtsev:2001,Newman:2001}.  In $1{,}000$ realizations, any node not in the giant SCC either had outgoing links only ($0.01 \pm 0.12$ nodes on average) or incoming links only ($0.02 \pm 0.13$ nodes).  Thus, the linear-chain structure of the SCC network (the color-coded SCCs in Fig.~\ref{fig_scc}) is not typical of a network with the same numbers of nodes and links.  One possible reason for this is that the in- and out-degree distributions of the combustion network are different from the Poisson distributions expected for the random counterpart (Fig.~\ref{fig_percolation}(a)).  However, even if we consider random networks with the same expected in- and out-degree distributions~\cite{Chung:2003} as the combustion network, the network still typically consists almost entirely of the giant SCC (having $11{,}067.96 \pm 6.01$ nodes on average, when estimated from $100$ realizations).  Moreover, the probability that a directed link is reciprocated is relatively high ($\approx 0.36$) in the combustion network, and this is far from typical for the random networks.  Indeed, for either of the two random models, the probability that a directed link is reciprocated is $p^2/[2p(1-p)] \approx 0.00061$, where $p = M/[N(N-1)]$ is the link probability.  These observations clearly indicate that the structure of the combustion network comes from additional constraints that the network must satisfy in order to represent Eq.~\eqref{cme}.

To see if the linear-chain structure is typical among the networks representing Eq.~\eqref{cme} (and not just for the \ce{H2}/\ce{O2} combustion process), we now consider the class of random chemical reaction networks defined for a given number $N_\text{r}$ of randomly chosen reactions involving a given set of $N_\text{s}$ species.  For simplicity, we consider only bi-species, bi-molecular reactions of the form $S_1 + S_2 \rightarrow S_3 + S_4$ in which the four species involved are distinct and chosen randomly from the $N_\text{s}$ species.  For each reaction chosen, we include the reverse reaction with probability $R$ (which can be tuned to reproduce the probability of reverse transition observed for the \ce{H2}/\ce{O2} combustion network).  For a given set of reactions and given the total number of molecules $N_\text{m}$ (which is conserved by any bi-species, bi-molecular reaction), we construct the corresponding matrix $\mathbf{W}$ through Eq.~\eqref{wij}, which in this case reduces to $W_{ij} = \frac{\kappa_n(T_j, p_j)}{N_{\mathrm{A}} V}\, m_{jk_1} m_{jk_2}$ for $i \neq j$, where reaction $n$ involves two reactants $S_{k_1}$ and $S_{k_2}$ and transforms state $j$ to state $i$.  Note that the factor $\frac{\kappa_n(T_j, p_j)}{N_{\mathrm{A}} V}$ can be omitted for this analysis as it does not affect the topological properties of the network.  For $N_\text{s}=8$, $N_\text{r}=33$, $N_\text{m} = 9$, and $R=0.39$ (leading to the probability of reverse transition $\approx 0.40$), we obtain networks with $N=11{,}440$, $M=156{,}576.60 \pm 1{,}760.33$ (yielding the average degree of $13.69 \pm 0.15$).  Despite accounting for the special structure of Eq.~\eqref{cme}, the network is still dominated by a single giant SCC, with an average size of $11{,}359.33 \pm 17.94$, representing $\approx 99.3$\% of the nodes (Fig.~\ref{fig_percolation}(b)).  Similarly to the purely random networks considered above, this class of random networks also undergoes a percolation transition close to average degree $=1$ (Fig.~\ref{fig_percolation}(c), blue curve).  The number of nodes with outgoing ($14.58 \pm 3.93$) or incoming ($33.62 \pm 5.12$) links only constitutes a tiny fraction of the network (though significantly larger than for fully random networks).  In particular, the nodes with only incoming links, which individually form single-node SCCs, corresponds to a tiny fraction of the states on which the steady-state probability concentrates.  This fraction becomes even smaller as the average degree increases (Fig.~\ref{fig_percolation}(c), orange curve).

\section{Conclusions}

The theory developed here reveals two distinct mechanisms through which non-monotonic dynamics emerge from non-normality of the network.  First, the non-normality moves the reactivity (the largest eigenvalue of $\mathbf{S}$) away from zero, enabling the decrease of the R\'enyi entropy and inducing non-monotonic entropy dynamics.  Second, non-normality allows the steady-state distribution $\chi_{0i}$ to be non-uniform and often highly localized (as we observed for the hydrogen combustion example in Sec.\ \ref{sec:non-monotonicity-condtion} and attributed to the DAG structure of the SCC network in Sec.\ \ref{sec:cca}), placing it off-center and close to a hyper-edge of the high-dimensional simplex of probability distributions.  This tends to widen the quadrants of growing deviation in the non-monotonicity condition we derived and thus increases the likelihood of observing an initial swing of the species concentrations away from their steady-state values.

These mechanisms for non-monotonic dynamics are intimately related to the fact that many chemical reactions, particularly in combustion processes, are irreversible or nearly irreversible.  In our formulation, reaction irreversibility translates to local directionality of the corresponding links in the network.  Since non-normality requires the network to be directed \cite{Asllani:2018,Johnson:2020,Hennequin:2012,Nishikawa:2006b,Ravoori:2011}, some of the reactions need to be nearly irreversible for the two mechanisms mentioned above to induce non-monotonic dynamics.  By removing the rare reverse transitions associated with nearly irreversible reactions, we reveal global directionality in the hydrogen combustion networks in the form of a linear chain of strongly connected components.  This directionality underlies the extreme localization of steady-state probability that promotes non-monotonic dynamics through the second mechanism above.

These two mechanisms for non-monotonicity requires non-normality, which in turn requires local link directionality.  However, our non-monotonicity condition clarifies that normal networks, even those with undirected links, can exhibit non-monotonic dynamics (but typically less pronounced than for non-normal networks), depending on the initial distributions of states and the choice of the observed quantity.  The observed quantities we considered here are similar to a weighted $1$-norm and thus fundamentally different from the $2$-norm of the deviation from the steady-state distribution, which can exhibit non-monotonicity only for non-normal networks.

The structure found in random networks is different (albeit still globally directional) and characterized by the dominance of a single giant connected component and even more extreme localization of the steady-state probability.  This holds true also for a class of random reaction networks that respects the network structural constraints imposed by the chemical master equation.  Thus, the emergence of a linear chain in hydrogen combustion networks must be due to other factors not captured by the random selection of reactions.  One possibility is that, when the product molecules are thermodynamically more stable than the reactants, this bias constrains the connected component structure.  While a model of random reaction networks accounting for this effect could explain the type of global directionality observed in hydrogen combustion networks, different factors could lead to different global structures in other types of chemical reaction networks, such as biological ones~\cite{Kaneko:2006}.  Many intracellular biochemical reaction networks that exhibit transient dynamics, such as signaling networks, are expected to have a high level of directionality.  The non-monotonicity condition we derived here and our SCC-based analysis of the network's global directionality are applicable beyond those networks we explicitly considered and lay a foundation for future research on non-normality and non-monotonic dynamics in complex reaction networks in general.  Our approach, which focuses on transient dynamics far from equilibrium and can be extended to externally driven systems using time-varying master equations, may extend to other network systems and also contribute to the ongoing development of non-equilibrium statistical mechanics.

\section*{Acknowledgements}
This work was supported by MURI Grant No. W911NF-14-1-0359 and ARO Grant No. W911NF-19-1-0383.

\section*{Appendix}
We prove that any right eigenvector of $\mathbf{W}$ associated with the (possible degenerate) zero eigenvalue can be chosen so that its components are all non-negative and sum to unity.  We also prove that the components are zero except for those corresponding to the nodes in the SCCs that have no outgoing links.  The latter implies that, for the steady-state distribution, all states outside these SCCs have $P_i > 0$.

Let $c$ denote the number of SCCs with no outgoing links and $c'$ the number of the other SCCs.  We first note that the DAG structure can be used to re-index the nodes and make $\mathbf{W}$ a block lower-triangular matrix, with each diagonal block corresponding to an SCC.  The matrix $\mathbf{W}$ then takes the following form:
\begin{equation}
\mathbf{W} = \left(\begin{array}{ccc|ccc}
\mathbf{W}'_1 & & \\[-6pt]
* & \ddots & \\
\rule[-9pt]{0pt}{6pt}* & * & \mathbf{W}'_{c'} & \\
\hline\rule{0pt}{14pt}
* & * & * & \mathbf{W}_{1} \\[-6pt]
* & * & * & & \ddots \\
* & * & * & & & \mathbf{W}_c \\
\end{array}\right),
\end{equation}
where an empty space indicates a zero block, and a star symbol indicates that the block can have non-zero components.  The $c$ blocks corresponding to the SCCs without outgoing links necessarily appear as the last ones, denoted by $\mathbf{W}_1, \ldots, \mathbf{W}_c$, which form a block diagonal submatrix in $\mathbf{W}$ because there cannot be links connecting these SCCs by definition.  Since the sum of components along each column of these blocks is zero, each block $\mathbf{W}_i$ has a zero eigenvalue along with the associated right eigenvectors.  Since each SCC is by definition strongly connected, and hence each $\mathbf{W}_i$ is irreducible, it follows from the Peron-Frobenius Theorem that the zero eigenvalue is not degenerate and that the corresponding eigenvector can be chosen to have only non-negative components.  We choose such an eigenvector, and we further normalize it, so the sum of the component equals one.  Extending these vectors to the size of the whole matrix $\mathbf{W}$ (setting all added components to zero), we obtain right eigenvectors associated with the zero eigenvalue of $\mathbf{W}$.

To see that the $c$-dimensional subspace spanned by these eigenvectors covers the entire eigenspace associated with the zero eigenvalue, consider a column vector of $\mathbf{W}$ intersecting with the diagonal block $\mathbf{W}'_i$, $i=1,\ldots,c'$.  The sum of the off-diagonal components of $\mathbf{W}'_i$ that appear in this vector is strictly less than the diagonal component, since the sum of all the vector components is zero and all the off-diagonal components are non-negative.  This implies that the Gershgorin circle corresponding to this column of $\mathbf{W}'_i$ is to the left of the origin of the complex plane and at a finite distance away from the origin.  Since this applies to all the columns of $\mathbf{W}'_i$ for any $i=1,\ldots,c'$, zero cannot be an eigenvalue of any of $\mathbf{W}'_1, \ldots,\mathbf{W}'_{c'}$.  Thus, all repetitions of the zero eigenvalue of $\mathbf{W}$ must come from the zero eigenvalues of $\mathbf{W}_1, \ldots,\mathbf{W}_c$, and the $c$-dimensional subspace constructed above is indeed the eigenspace of $\mathbf{W}$ corresponding to eigenvalue zero.  We note that the eigenvectors of $\mathbf{W}$ spanning the eigenspace were chosen to have components that are all non-negative and are zero outside the SCCs with no outgoing links.  Thus, any eigenvector of $\mathbf{W}$ associated with the zero eigenvalue shares the same property and can thus be normalized so that its components sum to unity, giving the steady-state distribution.  Consequently, we have $P_i>0$ in the corresponding steady-state distribution only for the nodes in the SCCs with no outgoing links.


\begin{thebibliography}{99}
\bibitem{foot3} The number $N$ of states accessible from the $2m$\,\ce{H2} initial state can be shown to be upper-bounded by a power law with an exponent of $8$.  To see this, we first note that any accessible state has the same number of \ce{H} and \ce{O} atoms as the initial state, which has $(4m+1)$ \ce{H} atoms and $(2m+1)$ \ce{O} atoms.  Here, we do not need to consider the \ce{Ar} atoms since the \ce{Ar} molecules do not change in the process.  Then, the maximum number of molecules of species \ce{H2}, \ce{O2}, \ce{H}, \ce{O}, \ce{OH}, \ce{H2O}, \ce{HO2}, and \ce{H2O2} that any state can have are $2m$, $m$, $4m+1$, $2m+1$, $2m+1$, $2m$, $m$, and $m$, respectively, implying that $N \le 2m \cdot m \cdot (4m+1) \cdot (2m+1) \cdot (2m+1) \cdot 2m \cdot m \cdot m = 4m^5(2m+1)^2(4m+1) \sim m^8$.  This bound, however, is typically far larger than $N$.  For example, $m=3$ yields $N = 347 \ll 4m^5(2m+1)^2(4m+1) = 619{,}164$.  A much tighter bound can be obtained by numerically enumerating all states preserving the atomic counts.  For $m=3$, this bound is $842$, as mentioned earlier in the text.
\bibitem{error_comment}
We find that the maximum error in $P_i(t)$ caused by the $\epsilon$-thresholding over all $i$ and over all $t$ between $t=10^{-8}$~sec and $t=10^2$~sec (well after the combustion process is considered complete) is small and generally decreases with the network size.
Already for the network of size $N=4{,}184$ built from the $16\,\ce{H2}$ initial state (the largest network for which the error could be computed), the maximum error is less than $2 \times 10^{-4}$.
\bibitem{foot1}Note that the Shannon and R\'{e}nyi entropies in this setting are distinct from the Boltzmann entropy, which includes uncertainty about the positions and velocities of the molecules in addition to the chemical composition. While the Shannon and R\'{e}nyi entropies may exhibit non-monotonic evolution, the Boltzmann entropy strictly increases according to the second law of thermodynamics.
\end{thebibliography}
\end{document}